# Bio-inspired Political Systems: Opening a Field

Nathalie Mezza-Garcia

Universidad del Rosario
La Candelaria, Bogotá, Colombia
meza.nathalie@ur.edu.co

**Abstract.** In this paper we highlight the scopes of engineering *bio-inspired political systems*: political systems based on the properties of life that self-organize the increasing complexity of human social systems. We describe bio-inspired political systems and conjecture about various ways to get to them – most notably, metaheuristics, modeling and simulation and complexified topologies. Bio-inspired political systems operate with nature-based dynamics, inspired on the knowledge that has been acquired about complexity from natural social systems and life. Bio-inspired political systems are presented as the best alternative for organizing human sociopolitical interactions as computation and microelectronics-based technology profoundly modify the ways in which humans decide. Therefore, weakening classical political systems. For instance, dwindling top-down power structures, modifying the notion of geographical spatiality and augmenting the political granularity. We also argue that, more than a new theoretical proposal, bio-inspired political systems are coming to be the political systems of the future.

## 1 Introduction

Human social systems are complexifying and it is becoming more difficult to frame and control them. At least, not through the traditional models of classical science. Bio-inspired political systems (BIPS) are an evolution of classical political systems. They are political systems based on the properties of life that self-organize the increasing complexity of the interactions among individuals and human social systems. In them, decision-making process follow metaheuristic algorithms inspired on nature, are tested via agent-based modeling and simulation, and are implemented by means of non-classical topologies. Bio-inspired political systems are an alternative to classical political systems because the latter are unable to handle the increasing complexity of the sociopolitical interactions in human social systems. We state, however, that more than an alternative, they will be, eventually, an emergence of the many shortcomings of classical political systems. Most of the limitations of the latter relate to the global view that classical political systems pretend to have of the systems they try to control. Classical political systems base their decisions in the false assumption that it is possible to have a perfect knowledge about all the behaviors, individuals, elements and interactions that conform human social systems or that play a role in political systems. Apart from that, they are convinced that in problem-understanding, decision-making processes and decision-implementation, linear structures and mechanisms are sufficient enough to reflect, handle or describe the mentioned interactions This belief leads classical political systems to institutionalize human sociopolitical interactions by means of political regimes with tree topologies.

The institutionalization of politics is what the Greeks called *politiké* [1]. It is a narrowed conception of politics because it is directly linked with the problem of governability, so it leaves besides many aspects of the public space which are not necessary institutionalized and form part, beyond institutions, of the political dimension of human social systems. Among them are ethical, philosophical, economic, administrative, religious, scientific, educational, aesthetical or social aspects. The *politéia,* the sum of the latter, is politics as a worldview [1]. It is where this paper stands for formulating the critique to the characteristics and properties of classical political systems –whether representative democracy, monarchies, dictatorships or others-, and classical political regimes. The critique starts from stating that the topologies of classical political regimes do not reflect the complex nature of the topologies of human sociopolitical dynamics and neither their decision-making processes evolve in accord with sociopolitical interactions. This is not a surprise: institutionalizing the complexity of human social systems entails reducing normalizing and standardizing it by means of linear topologies and decision-making processes. This makes classical political systems incompatible with organizing complexity and, even less, with harnessing it. Harnessing complexity means to "explore how the dynamism of a complex adaptive system can be used for productive ends –instead of eliminating complexity" [2]. This could be done by means of political systems with more organic topological aspects and decentralized decision-making dynamics -biologically motivated.

A political system is the aggregate of decision-making processes in a social system –human social system. Political regimes are the institutional scaffolding and rules of a political system. Decision-making processes, individuals, elements and relations in political systems are so diverse, vast and non-linearly interconnected that there is no reason for political systems to be expressed and planned in such a non-complex way through classical political regimes -and, even less, through their tree topologies. The institutionalization of politics is, however, the current state of things. Such state is being left behind as technology complexifies the means in which humans interact at an accelerated rhythm. For instance, providing easier, faster and cheaper ways to trade, communicate and travel -physically or virtually. The aim of this paper is to open the quest to engineering bio-inspired political systems, using the same platforms that are making the interactions within and among human social systems more complex and, at the same time, uncontrollable by top-down political structures. Namely, via computers, computation and, ultimately, microelectronics-based technological advances. The type of engineering we refer here is complex systems engineering [3]. The latter is interested in uncertainty, evolvability, adaptability, resilience, robustness, self-organization -among others-, instead of prediction, stability, reliability and centralized control [4].

Computation refers to information processing in computers (classical computation), computational systems (as Internet) and physical or biological systems

(natural computing). Probably the most common example of computation among human social systems is Internet. Since the invention and later popularization of computers, there have been many social changes and advances related to how the processing of information among human social systems has evolved. On one side, more powerful computation has allowed getting people closer despite geographical distances, it has helped get faster and more comfortably from one place to another, and has accelerated the propagation of ideologies and ideas among groups and societies. On the other side, it has helped gain knowledge about the complex world in which we live, studying phenomena and behaviors that used to be a mystery, misunderstood or unknown -which is what spearhead science and engineering are doing nowadays. For instance, bio-inspired algorithms used in metaheuristics have helped find better ways to solve problems, thanks to the creation of models that imitate how living systems develop, evolve and interact with their environment [5]; farther more, modeling and simulation has widely benefit the learning about computation in emergent dynamics, dynamics of self-organization and collective intelligence in living systems [6]; and, besides, the discovery of fractal geometry [7] and complex networks has conducted to recognizing and understanding more clearly some of nature´s structures and topological features. The three vias presented here (metaheuristics, agent-based modeling and simulation and complex topologies) are supported in the possibilities that computing provides. Computing is the main tool for engineering bio-inspired political systems, but it does not mean it should be the only one.

The claim that there is a tendency towards bio-inspired political systems and the description of the ways to get to them is studied into six sections. Firstly, the idea of bio-inspired models is contextualized. Secondly, bio-inspired political systems are introduced and some important remarks related to their design and engineering are marked as substantial. This leads to the study of the first, second and third order relationships between the three selected vias in which bio-inspired political systems can be engineered. Fourthly, some background elements of bio-inspired political systems are mentioned, showing how classical political systems and their structures are being affected, giving rise to a tendency towards more organic political systems. Fifthly, some possible implications of bio-inspired political systems are grasped. Finally, the paper concludes with several important remarks related to engineering bio-inspired political systems.

## 2    The Shift towards Bio-Inspired Models

A model is an abstraction (simplification) created to understand a system or phenomenon. Models should be as similar in structure to the *objects,* phenomena, behaviors, systems or problems that are being modeled [8]. For a long time, classical physics was the base for models in science –even political science. The result was simplistic models focused on analysis, control, predictability, rigidness, determinism, stability, equilibrium, certainty, centrality, reliability. That is, linearity. Classical political systems –and classical political regimes- are examples of physics-based models in political science because they strongly focus on the properties mentioned above. They both are reductionist approximations to the complexity of human social systems. This makes them to be designed with hierarchical centralized control mechanisms, cause-effect dynamics and top-down imposed normativity.

Figure 1 shows a model for political systems shared by the mainstream of political science. It was developed by Jean-William Lapierre [9], based on David Easton´s model (see [10]). One of its many shortcomings is that it is conceived as a deterministic cause-effect system, where dynamics are understood as a linear sum of decision-making processes. With no doubt, we can claim that the model could have based on the theoretical implications of Newton´s laws of motion. In general terms, the model errs in trying to schematize human sociopolitical dynamics from a non-complex point of view. A possible reason for this can be that the model was developed following a *general systems theory* perspective [9]. Correspondingly, it assumes a perfect knowledge about all the parts and interactions involved in the decision-making processes of political systems.

It is not a secret that if we are referring to complex systems not even knowing all the elements and interactions we can talk about determinism or perfect knowledge. When referring to political systems, this is also impossible. Political systems are imposed over human social systems and the interactions in them involve such complexity that not even when institutionalizing them by linear mechanisms trough classical regimes their complexity is eliminated. Maybe the farthest that this analytical and systemic models have reached has been to cooptate some of the traditional tools of the sciences of decision such as system dynamics, decision trees, real options and portfolio management [11]. Anyhow, understanding the *black box* of political systems assuming linear relations should never be tried to be done again.

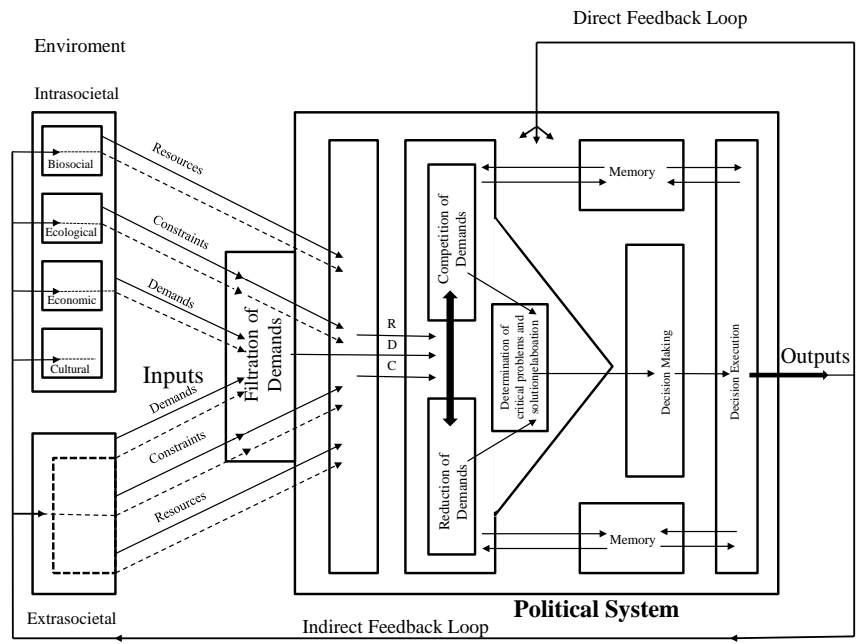

Fig. 1. Model of political systems developed by Jean-William Lapierre [9]

Instead of physical-based models, and in contrast to the deterministic world showed above, human social systems -the systems that political systems try to organize trough political regimes- are feasible to be described by properties much closer to biology, such as evolvability, adaptation, uncertainty, emergence, self-organization, learning and synthesis. The paradigm of complexity are biological systems. More precisely, complexity has life as its core. For this reason, these properties are widely studied by the sciences of complexity. Therefore, for understanding and trying to organize human social systems, the best alternative is turning to complexity sciences. Notably, to bio-inspired models. This would benefit of the fact that complex systems engineering is interested at the organic properties mentioned above for generating close to real life models too. Complex systems engineering recognizes that bio-inspired engineering is a way to show how engineering is complexifying [4] –as our world complexifies as well. This, a union between political systems, bio-inspired models and complex systems engineering sustains in these motives.

The sciences of complexity have developed models, theories, concepts and tools for approaching non-linear –complex- behaviors, phenomena or systems. A great part of the most recognized models that complexity works with are more organic than those of classical science –without this meaning that they are more complicated. This allows a better comprehension about the systems that exhibit life-like behaviors. Among them, human social systems. We claim that the apprehension of complexity in political systems and political regimes is the best way (maybe the only one) to organize and harness the increasing complexity of human social systems and their interactions. It entails engineering bio-inspired models that replace classical ones and that reflect better the structures of human sociopolitical dynamics than those referenced the most in the study of politics and the political.

Every discipline among the human and social science studies complex systems. Many of them have already apprehended complexity and have given a certain shift towards bio-inspired models, although not necessary their mainstream. Economy recognized the chaotic, non-linear and self-similar nature of markets behaviors [12]; sociology acknowledged the complex adaptive nature of human social systems [13]; history recognized that it is not a sum of facts from the past, but a non-linear and open system that considers even facts that never happened [14] and, finally, the algorithmic complexity of art was recognized, enabling to think about the scopes of measures for the complexity of artistic pieces and their relation with subjective experiences [15]

The most recent models they now use have life as the ground for building their explanations. In political science, however, this is not the case. Political Science studies some of the most complex systems that exist on earth –individuals, groups and human social systems- and, despite it, it is one of the disciplines among the human and social sciences that when working with complex systems has not completely recognized complexity. Hence its attachment to the classical realm of science. For approaching complex phenomena, classical models are non-viable anymore. In fact, they have largely demonstrated a wide spectrum of limitations for times of increasing complexity [16].

Figure 2 points to how political systems need be complexified turning to biologically-inspired models that present life-like behaviors –in this case, for understanding better the behavior of complex systems, finding better solutions to complex problems and designing political structures that reflect complexity. The reason is that life is the phenomenon that (i) presents most complexity, (ii) harmonically manages to self-organize, and (iii) harness complexity the most.

Bio-inspired models do not necessary have to be complex models. They can be very simple and still be complexity-based. The importance relies on the complexity of the dynamics they describe. In any case, it is important to bring up that the quality of life of a human social system largely depends on the complexity of its political system. Complex models are the best known road that can be selected for thinking about models for political systems -particularly bio-inspired models. As it will be shown, an advantage of the latter is that in the sciences of complexity, the latter are computational.

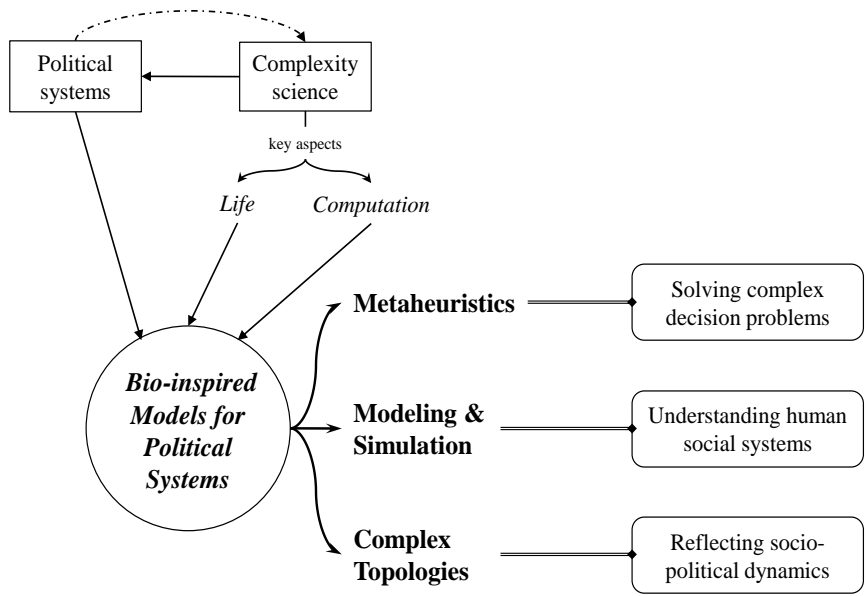

Fig. 2. Bio-inspired models

As explained in [17], the interest on life has been addressed from early philosophers to contemporary scientists, either interested in describing life, the nature of life or life´s hallmarks. Originally, life was seen as a binary state: *something* was either alive or dead. Some contemporary approaches center their attention on whether the difference between the living and the non-living relies on composition, structure, function or a combination of the three. Ultimately, the differences between the living and the non-living are qualitative, in terms of degrees and of organization [18]. Computational models that study life-like behaviors have greatly contributed to this conception because in the middle of the living and the non-living there are computational bio-inspired models that have life-like behaviors. Additionally, they have taught us that the living properties of single individuals can be extrapolated to their social systems and, from a Darwinian point of view, we can state, for instance, that populations of human individuals -human social systems- present life-like behaviors. Therefore, it is valid to study them by using bio-inspired models that describe their life-like dynamics.

## 3   Engineering Bio-Inspired Political

Engineering *bio-inspired political systems* (BIPS) implies to design *non-classical* and self-organized political systems where:

a) Decisions are the result of metaheuristics processes.
b) Comprehension and explication of sociopolitical phenomena are the result of (agent-based) modeling and simulation.
c) Bio-inspired topologies of political regimes are the result of complex network structures.

Engineering these types of models is important because political systems face problems about human social systems. However, we must remember that in many situations, the problems involve humans but also other species in the planet or the biosphere itself. As a result, in some cases the scopes of the decisions taken by political systems entail negative bioethical, social, economic and political consequences. One explanation to this is that most of the time the decisions that are going to be implemented are not previously tested; the systems upon which political systems impose their decisions and environments are not well comprehended; and there is not a correspondence between the complexity of the affected systems and the linearity of the methods of decision-implementation. The best way to overcome these shortcomings is by approaching towards the engineering of models much closer to the complex nature of the systems affected by political systems. Bio-inspired models come as a substitute of classical models for political systems and regimes because there are some costs for maintaining the hegemony of them (lives, extinctions, economic or social consequences…) that should never be assumed anymore by any individual, species or population in the planet.

We decided to focus only in three ways for engineering BIPS: metaheuristics, modeling and simulation, and complex topologies. Their utility relies on their practical, theoretical and conceptual relevance, but it does not mean that new *teqniques* could not be incorporated to their logic in the comming years. The three vias mentioned above, when taken separately, are first-order models. The interaction between two of them, in any direction, (a ∧ b, a ∧ c, b ∧ c) form second-order models, which we named basic hybrid models. And the interplay between the three –or more- conform third-order models and are the long-term desired scenario we think is needed for letting human social systems self-organize without the need of any imposed or elected ruler; or top-down system. Figure 3 shows the possible interactions of first-order, second-order and third-order bio-inspired models. In the following paragraphs we explain with more detail the role of each road for engineering BIPS and at the end of the section the possible relationships among them.

### 3.1 Using Metaheuristics for Decision-Making Processes

Political systems face problems by taking decisions upon phenomena that concern various kinds of complex systems, apart from human social systems. Consequently, the problems that political systems try to solve are complex problems. Metaheuristics are a tool for solving complex problems. They are crucial in bio-inspired political systems because of the complexity that characterizes the systems upon which they are imposed. Certainly, human social systems require better methods for problem-solving than those provided today by mere intuition of governors and based on analytic and reductionist methods.

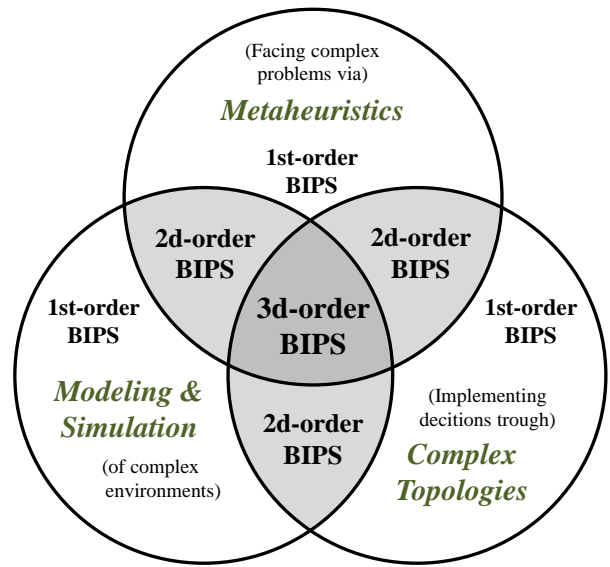

Fig. 3. Bio-inspired political systems: first, second and third order relationships

Indeed, because "the main task of management [in political systems] consists of optimal decision-making" [19], the problems that political systems try to solve can be understood as optimization problems. I.e. problems that look for finding the best decision (which is not necessary the optimal) in a given moment.

Given the non-trivial nature of the problems, they should not be tried to be solved by decisions taken without rigor -theoretical, conceptual or ethical. However, decisions in political systems depend on the decision capacities of governors and the traditional methods they usually work with. That is, their own personal interests - which are not always oriented towards the general welfare- and limited tools of classical science. Theoretically, *they being there* means that they could solve and find the best alternative to any problem they face, thanks to the *unique* and in anyway incipient swarm-like collective intelligence of their voting and deliberation processes. In reality, having top-down methods for problem facing and individuals with their personal interests for deciding upon a global view of a system is not enough. Ergo the task and the role of governors are actually naive. This does not mean that complex problems in human social systems are always going to be faced by without rigor. Notably, the use of metaheuristics can help finding solutions close to the optimal.

Metaheuristics are computational tools for resolving complex problems regarding optimization and prediction -problems that cannot be elucidated by traditional analytical methods. Metaheuristics can be described as "general-purpose algorithms that can be applied to solve almost any optimization problem" [5]. They allow optimization under uncertainty contexts.

There are some metaheuristics that are physic-based, mathematics-based, biology-based and ethology-based [20]. The ones in which we are interested in are population-based and biologically motivated. They use algorithms inspired on natural phenomena or in the way in which some species solve problems, translating the processes into general frames that can be used for modeling various kind of complex phenomena. Metaheuristics start from considering that for any problem there is a defined space of multiple solutions. Population-based metaheuristics randomly search in the space of solutions and combines the best solutions between them, so in each generation the robustness of solution increases. It can be said that despite the different bio-inspired algorithms in metaheuristics, the basic metaphor is evolutionary.

This ways of finding solutions to problems is steps beyond how political systems do it nowadays. The following is a list with the main families of metaheuristics.

- Neural computation
- Evolutionary computation
- Swarm imtelligence
- Inmune computing
- Membrane computing

The reason why complex problems should be faced by complex tools is because there are problems that (a) can have infinite solutions, (b) have dynamic solutions that exist in time-changing environments (c) are constrained and obey to restrictions, and (d) are based on contradictory principles because there can be many possibly conflicting objectives [21]. In other words: first, there are always missing pieces for the puzzle; second, there are pieces of the puzzle that fit in an x time, but not in a y time; and third, if a piece fits (in an x time), other pieces disengage (in the same x time o in a y time) [22]. These are complex problems. In them, finding *a* solution, *the* ultimate answer is impossible, or it would take millions of years if not an infinite computational time to be calculated. That is why the answer to life, the universe and everything is not 42[1].

As metaheuristics will become more used by engineers and decision-makers [5], they should be promoted as one of the best tools there are for optimizing complex problems. We claim that in the future the magic of decision-makers will rely on whether they know how to translate a problem of political systems in terms of metaheuristics optimization.

### 3.2 Agent-Based Modeling and Simulating Bio-inspired Political Systems

"Agent-based modeling is a recent simulation modeling technique that consists on modeling a system from the bottom-up, capturing the interactions taking place between the system´s constituent units" [23]. Agent-based modeling (ABMS) was born in the context of artificial life (AL), which creates synthetic life on computers that exhibit life-like properties and behaviors [24]. For the process of modeling, the system is understood as a collection of components (agents, parcels) that nonlinearly interact and give rise to emergent patterns and behaviors that cannot be directly traced back, simply, to the properties of the parts taken separately. ABMS can be for specific or general uses and can have strategic, tactical or operational domains [25]. By defining a set of basic rules, we can observe how patterns start to emerge bottom-up. By viewing the evolution of a system along generations, we can have deep insights to the comprehension of complex systems. This is something that classical modeling techniques lack because they fail in being able to work with nonlinearity [25].

---

[1] Here we refer to the science fiction movie directed by Garth Jennings, The Hitchhiker's Guide to the Galaxy, where, in an ironic manner, a computer built by pan-dimensional beings calculates 42 as the answer to life, the universe and everything.

In sum, according to [6], "in agent-based modeling, a system is modeled as a collection of autonomous decision-making entities called agents. Each agent individually assesses its situation and makes decisions on the basis of a set of rules". Agent-based modeling and simulation (ABMS) becomes a useful mindset when [6]:

- Agents exhibit complex behavior
- The interactions between the agents are nonlinear, discontinuous, or discrete.
- The topology of the interactions is heterogeneous and complex.
- The population is heterogeneous and each individual is (potentially) different
- Space is crucial and agents´ positions are not fixed.

Political systems can benefit of ABMS because they are expressed in organizations and institutions -which are "often subject of operational risk [6]", and organizations are one area of application of ABMS. In this light, ABMS can help political systems by understanding the complexity of the sociopolitical dynamics having place in and upon them. But it can also help with internal (organizational) -sometimes topological simulation. Thereby, lowering the impact of the operational risk of political systems due to the valuable information that ABMS provides about the behaviors of modeled agents and decisions in complex systems.

Political systems have always implemented –and are implementing, still- sets of decisions taken without even proving if they are good solutions. As a result, many times the decisions that are supposed to be favorable for a population result, instead, in negative outcomes and effects. This dues to the fact that nor the problem or the social system are fully comprehended. So decisions, in most cases, are arbitrarily imposed. For this we can say that almost every decision that classical political systems impose upon human and natural social systems are experiments with the real world, where governors are the scientists and the world is the laboratory.

Not testing the decisions that will further on be executed usually takes to two types of negative outcomes. Fist, aspects related to time or treasury and, second, and more importantly, bioethical results, such as the loss of lives, killing bio-diversity, augmenting poverty or polarizing more the world. With ABMS many of these negative outcomes can be avoided because of the gained comprehension about the systems that concern decisions and where decisions are implemented. ABMS is the best alternative so far for creating simulated environments that help testing solutions and decisions without having real-life individuals are guinea pigs.

The scopes of ABMS will bring questions about the role of future politicians as decision makers if ABMS continues to become more proficient at decision-making processes than governors. However, while this fully occurs, ABMS should be more used by public decision-makers, helping to anticipate potential outcomes or implementing –better informed- decisions [25]. Experimental proves in artificial life help narrow error margins when implementing a model or solution in real life.

In an on-going research using agent-based simulations, we manage to synthetize self-organized control mechanisms that adapt over time with changes in the environment, using only local information. Our quest is to find out whether human sociopolitical interactions, when they are not mediated by institutionalism, succeed on making coordinated patterns to emerge. Everything indicates that when defining basic elements in the base of the social system, it is plausible that human social systems give rise to adaptive and intelligent collective *swarm-like* behaviors. Unquestionably, ABMS is the best way to gaining comprehension about emergent behaviors in complex systems. Political systems need to appropriate of their use[2].

### 3.3 Thinking Complex Topologies for Political Regimes

Topologies refer to the distribution of nodes in a network. Tree topologies (figure 4) are the structural models for institutions in classical political systems. That is, for classical political regimes. Among the classical topologies presented in the figure 4 tree topologies are the ones that represent complexity the least. They are suitable for imposing restrictions to complexity because they have centralized control executed by means of a node in the top of the structure. This node *is* aware of all the information going throughout the system and it can be an emperor, a king, a queen, a prime minister, a president, a dictator, a parliament or a congress. Basically, any individual or group in charge of the direction of the decision-making processes in classical political systems.

Tree topologies for political regimes are obsolete for times of increasing complexity. It is not plausible for a single node in the top of a political structure to continue trying to have global information about all the dynamics of the complex systems over which it imposes upon. Therefore, organizing the complexity of human social systems should not be done by top-down methods, but bottom-up synthesis. In that way, complexity can be organized better and can be harnessed too. Nevertheless and despite that this is actually how sociopolitical interactions occur in human social systems (by bottom-up synthesis), the mainstream of science has been permeated with the idea that sociopolitical interactions must be top-down controlled.

Bio-inspired political systems recognize the importance of the topological properties of interactions for computational purposes, being some topologies more suitable for better information processing than others. We claim that the topologies of political regimes need be based on models of complexity because that is the nature of the systems they organize. A correspondence between the physical operations of the structures of political regimes and the logical structure of human social systems is needed, for them to reflect the computational structures (information processing) in human social systems. That is, they need to be isomorph or, even better, merge with sociopolitical interactions.

---

[2] Most of the cases where ABMS has been used in Political systems have been for activities related to military and war purposes.

Figure 4 shows various kinds of models for topologies: classical, hybrid and complex network topologies. Among the classical models, we present basic structures for tree, bus, star, mesh, fully connected and ring topologies. We included within hybrid models those topologies with fractal structures and random ones, conformed either by single nodes stochastically distributed or models formed by other classical topologies, but different from tree topologies.

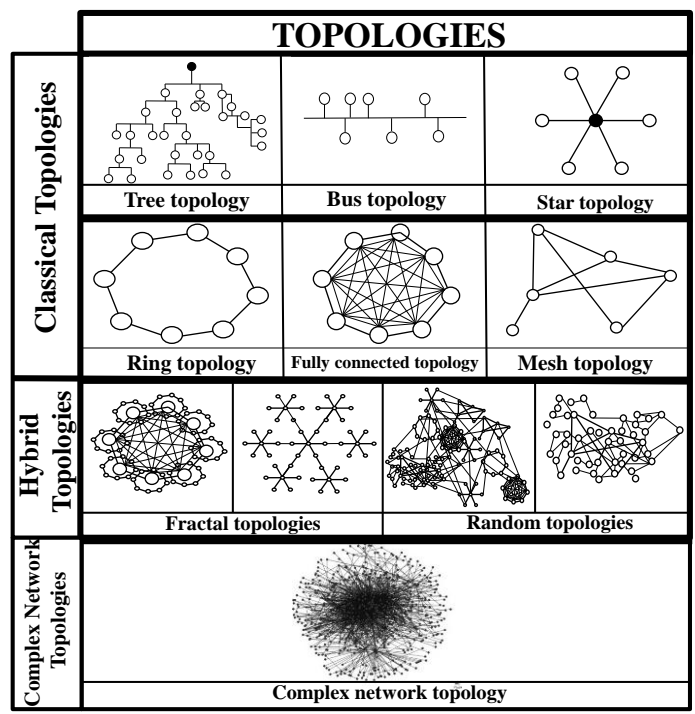

Fig. 4. Classical, hybrid and complex topologies

Figure 5 characterizes topologies. It has three axes. In one extreme of axis Z we located the property of being physical-based for topologies models and in the other extreme those more biologically motivated. In this case, tree topologies are characterized as the most physically-based, whereas complex network structures are presented as the more biologically-motivated. Axis X corresponds to how centralized or decentralized a topology is. Again, tree topologies, in this case, together with star topologies, are in one extreme –the most centralized. Axis Y goes from the most linear to the most complex. For this case, complex network structures are situated as the most complex of all and tree topologies are among the most linear. Most of the times, in the middle of the axes we located hybrid models and the rest of classical topologies. Even the latter are preferable than tree topologies for the structuration of the regimes of political systems because, despite that their natural emergence is highly improbable, they are more decentralized than tree models.

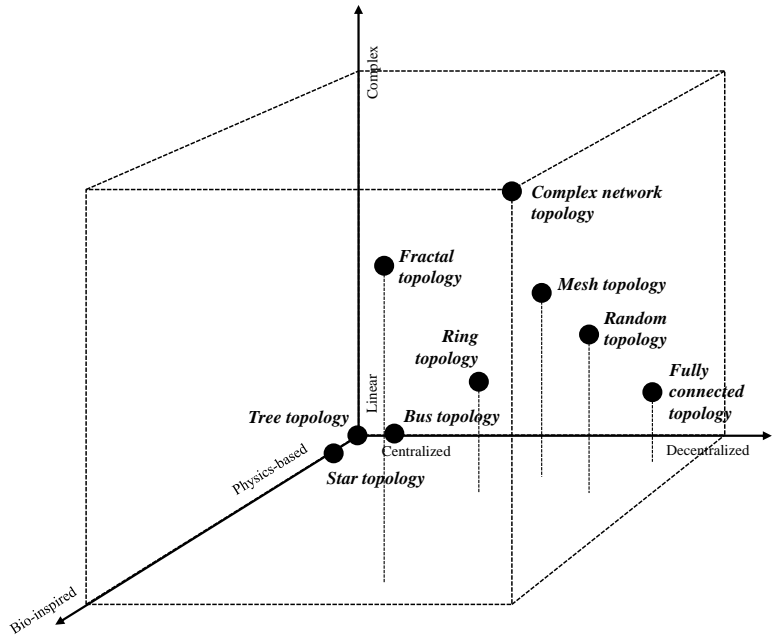

Fig. 5. Topology characterization

The structures of the regimes in political systems must reflect the structures of the complex sociopolitical dynamics over which they are imposed upon. Complex political topologies are key in bio-inspired political systems because they are the result of sociopolitical dynamics synthetized bottom-up, which imply a better organization of the interactions between individuals, human social systems, among them and with their environments.

### 3.4 Second and Third-Order Bio-inspired Models

As figure 3 shows, second-order models correspond to the interactions between two of the three first-order models: (a) population-based metaheuristics, (b) agent-based modeling and simulation, (c) complex models for political topologies. That is, we can combine metaheuristics and modeling and simulation, metaheuristics and bio-inspired topologies or bio-inspired topologies and modeling and simulation. However, there will be times in which one might prevail over the other, which means that for second-order models, there are actually six combinations instead of three, as following.

1) (a / b): When facing problems of optimization, the result of population-based metaheuristics can be tested in simulated environments before implementing them in the real world.

2) (b / a): Agent-based models can be enriched by metaheuristics and, in that way, we can have models much closer to reality.
3) (c / a): The more complex a topology is, the more it becomes a favorable environment for being receptive towards solutions and the logic itself of metaheuristics coming from non-mainframe power concentrators –or governors.
4) (a / c): Metaheuristics can serve as a parameter for designing and deciding which topology to implement in which case for organizing certain sociopolitical dynamics.
5) (b / c): The local information of ABMS dynamics would reinforce the processes of looking for topologies in congruence with the complexity of human social systems. Modeling and simulation could help the morphogenesis of political systems, finding topologies that actually reflect the structures of the sociopolitical dynamics over which they are imposed. It serves for seeing and proving the functioning of each topology.
6) (c / b): Decentralized topologies are suitable spaces where the results of agent-based modeling can be taken into account because there would not be a central control deciding upon how a system behaves.

On the other hand, third-order models are relationships between the interactions of three –or more- models. Third-order bio-inspired models are the most plausible way for avoiding the deviations that occur when politics is institutionalized. They would be fully self-organized sociopolitical interactions with synthetized adaptive and self-organized bottom-up *control*. However, promoting self-organized models for organizing social systems in a world that has always been controlled top-down would imply non self-organized ways or anarchic spaces for it to occur.

Non self-organized vias for changing a political system or regime mean violent mediums, as the stories of most of political revolutions have proven. The other extreme, although pacific, would take too long. We would have to wait until the top-down structures of classical political systems continue to progressively weaken by the effects of technological advances in humans´ exchanges of information. The price to pay would be that it will be extremely late for the wellbeing of some human groups, their habitats and for other species that inhabit them, and that are currently been affected by the decisions taken in the structures of classical political systems.

A good solution to accelerate the existence of third-order models would be to find an intermediate state between self-organization and design. Self-organizatin could be induced by means of a guided self-organization. I.e., engineering systems that tend to self-organize their dynamics. The, apparently, contradiction between both methods (design as planning and self-organization as non-determinism) was solved by Prokopenko [26], who found that combined both could lead to a point where self-organization is at the base of the desired (designed) dynamics. In the case of bio-inspired political systems, using metaheuristics for facing complex problems, modeling and simulating the enviroments where those problems take place and having complex topologies were decisions are bottom-up implemented would be the parameters that *guide* the self-organized dynamics of the system.

## 4 The Complexification of Human Social Systems: Bio-Inspired Political Systems´ Background

Bio-inspired political systems will be an emergence of the interconnection of some phenomena that are non-linearly transforming human social systems. Independently of the engineering of BIPS, there is, indeed, an increasing tendency from classical political systems towards more organic ones. The reason relates to some phenomena that are occurring in contemporary world and that are deeply transforming political systems by moving them away from how they have always been. Some examples are: the weakening of traditional power structures in top-down political systems, the reinvention of the local as the confluence centers for sociopolitical interactions or the propagation of ideas crossing artificial national boundaries. These facts serve as background for supporting the idea that, as the interactions between human, natural and artificial system get complexified, there is a tendency towards bio-inspired political systems. The following are some phenomena that, when combined, reinforce the tendency toward BIPS and the need of a field for their study.

**4.1 The Network Society**

As technology (microelectronics-based) complexifies the means trough which humans communicate -Internet, social networks, transportation, mobile phones, computers, and others, human communication modifies with it [27, 28]. This fact is in great part responsible for the small world phenomenon [29] that human social systems present. It is also the base of the concept of the network society [30], developed by Manuel Castells, according to which is "the social structure resulting from the interaction between the new technological paradigm and social organization at large" [31].

Network societies can be considered as an input for engineering BIPS because this phenomenon is decentralizing human interactions. Bounds are becoming more flexible and adaptive and are basing more on coordination than imposition. Of course, the network society implies technological basis for the communications of individuals, but not every individual –or political grain- in the world has access to them. However, we recognize that, along human history, the life conditions of individuals with less economic capacities has been increasing [32]. We hope that the same happens with the access to technology in the very long term.

There is something about sociopolitical relations in the network society that BIPS may influence in a positive way: "[the network society] excludes most of humankind, although all of humankind is affected by its logic and by the power relationships that interact in the global networks of social organization" [29].

This is supported by the fact that the network society brings up the phenomenon of *networked individualism* [31], which consists on how individuality becomes the center of social structures. Starting from there, individuals express their individuality by sociopolitical interactions facilitated by the connectivity in the network society. This leads towards public political spaces mediated by virtual interactions because technology provides more active and direct means for humans to connect to each other, to have voice, presence and action in the emerging non-geographical political arena [33, 34]. Thus, sociopolitical interactions facilitated by technology aim to work with less artificial geographic, migration, territorial, economic, ethnic, cultural, political or social restrictions. These types of societies claim for new political structures, better decision-making problems and better understanding of their problems.

**4.2 Deterritorialization of Space**

A great deal of the interactions taking place today in human social systems occur in non-geographical spaces; That is, by means of the Internet. An appropriate word is that they are *deterritorializing*, which means that every time it is becoming more difficult to link personal identities with a defined geographical territory or population. Identities are becoming more complex networks of experiences than clusters of traditions. This is explained by the fact that particular data about human social systems (religions, costumes, traditions, ideologies, fashions, styles, hobbies, etc.) is spreading more rapidly among them. Internet is a platform by means of which individuals can have access to *worlds beyond their own*. In that way, they can look for spaces and activities more related to their personal *wished* identities than to the geographical territory where they were born in. Two precisions must be brought out. First, complete deterritorialization will not happen fully unless Internet becomes affordable and accessible for underdeveloped and later-developed countries and not only for the so called developed nations. And, second, despite all the advantages, deterritorialization must not be a condition that should be pursued because it has many non-desired implications, for example, for the historic memory. Nevertheless, not pursuing it is not happening.

Deterritorialization comes after exponentially augmented information flows among human social systems. It implies that the group marks that used to facilitate tagging a population with certain symbols, shared features, generalized personal identities are no longer possible to be generalized. As a consequence, it becomes less easy to control human social systems by means of tree topologies because there are no generalized patterns that can serve as frames, such as adscription factors, symbols, anthems, flags, etc. or belonging to a defined geographical space. The relation between deterritorialization and bio-inspired political systems is that as the interactions in human social systems become more difficult to control by means of top-down mechanisms, they tend to self-organize and form complex sociopolitical networks or, at least, hybrid BIPS.

**4.3 Finer Political Granularity**

The political granularity measures the extension of the territorial parcels over which a political regime imposes its modes of organization upon the social systems that remain in a specific territory. The term *granularity* was extrapolated from molecular dynamics and engineering [35, 36]. For practical purposes, the parcels are called grains and their extention defines their granularity. The States' territories, provinces, regions, states and cities are examples of grains. Figure 5 sumarizes three momments of the story of political parcelss: empires, kingdoms and modern states. Every one of them exceeds its predecesor in energy consumption. The graphic shows how since the history of big empires, energy consumption has being in augment and, as this happens, the extensions of the territories over which the institutions of political systems impose their mandate have been decreasing. Thus, we can say that there is a non-linear tendency of progressively finer administrative and territorial political granularity going from coarse-grained parcels to fine-grained parcels.

Our *longue durée* approach [37] suggests that this counts even for the cases where political coarse grains are formed, at first, from political finer grained parcels; i.e. the EU conformed by states or big empires formed by smaller territorial organizations. Although it is possible that in the future some coarse-grained parcels that reunite finer-grained parcels will continue to emerge [38], it is higly umprable that they will continue to have the control capacities they still have today due to the possibility of using physical violence –the basis of past and contemporary political power. This is explained by the idea that as human interactions will no longer only occur in a space with a geographical conception of territory, it will become more difficult to apply coercion over a population or territory, as it is becoming more difficult to frame them. In addition, parceling the globe´s territory and management in finer-grained political parcels, reduces the capacities of mainframe power concentrators.

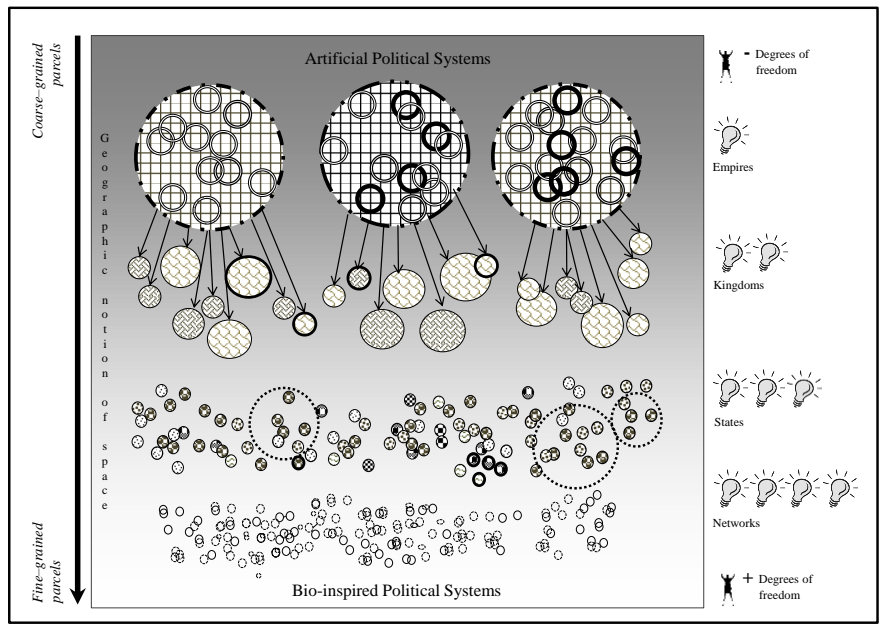

Fig. 6. The evolution of political granularity

Finer political granularities lead towards bio-inspired political systems because as political grains become smaller, the importance of local information increases. Many contemporary examples of decentralized cooperation and trading among cities or between cities and national governments are proven to be highly effective, in comparison to nation-nation trading and cooperation [39]. Cities are political grains as well, and they cannot continue to be ruled by political institutions that operate in non-local levels, missing important information about the complex micro and local dynamics of the system.

When focusing on a geographical notion of space, finer political granularities mean non-centralized control. And, when it comes to non-geographical approaches for understanding territorial space, the idea of control, anyway, starts to vanish. Then, finer political granularities conduct towards BIPS, which are not control-focused. So the evolution towards finer political granularity leads to bio-inspired political systems. The link between the size of the grains and energy consumption shown in graphic 6 suggests that there is a propensity towards finer political grains –formed by complex network structures.

### 4.4 Sociopolitical Self-Organization

Sociopolitical self-organization consists on how political interactions among individuals bottom-up synthesize. This occurs without the need of any leaders, governors or top-down regimes. Sociopolitical self-organization is the *politéia* in its broadest sense. The contemporary expression of the phenomenon is being facilitated by the diffusion of technological means humans use to trade, travel or exchange, bypassing intermediaries, such as classical political systems and their regimes. Sociopolitical self-organization, besides being a continuously adaptive emergence of BIPS, as we previously stated, is also an antecessor of them. To the extent that humans can self-organize their sociopolitical interactions, hierarchical means and control are being left behind [27]. Together with this, the increasing diversity of human social systems, sociopolitical self-organization is pushing classical political systems to try to be more inclusive by becoming more open to diversity, which is one of the symptoms of a society that will collapse [40].

For instance, every time there are laws that try to regulate social aspects of individual´s life, such as their sexual orientation or tendencies. With this, the hegemony of *unique* ways of thinking, living and manifesting individual identities are becoming a remembrance of the past. Regulating and legislating about such increased diversity in individuals personal and group identities by means of top-down structures is becoming more difficult -even more when individuals are recognizing their capacity to self-organize using social networks, not needing political intermediaries. Social changes are occurring at a rate so fast that when the formalisms in political systems are just starting to consider them, many new ones have already emerged.

In respect to social networks, the revolutions, insurgencies, marches and rebellion that are self-organized by using them are proving that the top-down structures of classical political systems cannot handle a world with increased complexity. Recent years provide many examples of sociopolitical self-organized dynamics, such as the Indignant Movement, the students' movement in South America and the Arab Spring [41]. The latter are emergencies that rose up against top-down political, economic and educative structures and regimes of classical political systems. There is nothing that points to the latter as contingencies. Rather, they must be considered as a tendency, as classical political systems continue to weaken and become openly incompatible with organizing and harnessing complexity.

Sociopolitical self-organization is also making political frontiers to become less rigid and more permeable. No longer a phenomenon taking place in a defined geographical space stays *in* there. It diffuses when it enters non geographical spaces (Internet), and afterwards it spreads again in geographical spaces. The international system is a space that feeds from self-organized geographical and non-geographical sociopolitical self-organized interactions. It is conformed by complex networks of international treaties and agreements, some of them voluntary and non-coercive, generated by diverse actors interested in cooperation, coordination, consensus and protection. The regimes act at micro or mezzo scales, such as cities, regions or states, but there are no generalized top-down regimes that apply globally for all the international system. The macro scale of it is increasing in importance and, as mentioned above, its contemporary basic inputs come from self-organized dynamics.

## 5 Implications of Engineering Bio-Inspired Political Systems

A world where bio-inspired political systems have already been engineered would differ in many aspects from the world in which we currently live. Its economic structures would not be sustained by exploitation, being part of a collectiveness would not be defined by the adscription to a geographical territory, political regimes would not be something that is already set up before someone is born and to which individuals must be subjected, the *politéia* would not have to be subordinated to the *politiké* and, certainly, there would be better possibilities and hope for biodiversity on earth –at least for what remains of it. All these, however, require, first, for instance, a more generalized diffusion of means such as Internet –or its successors- along the world. We now turn to describing some of the implication of BIPS. We must clarify that the higher the order of the logic of the interactions between the vias for engineering BIPS, the more radical the implications mentioned here. For them to fully occur we have to wait some generations.

**5.1 A Complex World has Non-Imposed Economic Structures**

Through their regimes, political systems try to organize or, at least, to influence economy in political grains. They do so by means of laws, legislation, public policies, international treaties and agreements. Non-imposed economic structures will emerge, in the absence of legislation that regulate economic exchanges. The outstanding economist of the twentieth century, Friedrich Hayek, envisioned this spontaneous order [42] in economic catalaxies. Catalaxies are emergent economic structures which are the result of self-organized process of specialization and exchange among individuals in human social systems.

Internet will play a remarkable role in the emergence of no-imposed economic structures because it facilitates trading activities. Indeed, currently, internet is eliminating the need of having many intermediaries for trading and exchanging by helping individuals get closer to a more independent economic status [32], reducing the scenarios where they could be exploited. At the same time, this entails better bioethical labor conditions than those provided, imposed, guaranteed or searched today by political regimes. Online crowdsourcing platforms and web pages where inventors can sell directly their work and ideas are examples that reaffirm the tendency [32]. The network society also plays an important role in this equation because the ability to work autonomously and be an active component of a network becomes paramount in the new economy: self-programmable economy. The same author who defined the concept also states [31]:

> [L]arge corporations decentralize themselves as networks of semi-autonomous units; small and medium firms for business networks, keeping their autonomy and flexibility, while making possible to put together resources to attain critical mass small and medium business networks become providers and subcontractors to a variety of large corporations; large corporations and their ancillary networks engage in strategic partnerships on various projects concerning products, processes, markets, functions, resources, each one of this project being specific, and this building a specific network around such a project, so that at the end of the project, the network dissolves and its components form other networks around other projects. This, at any given point in time, economic activity is performed by networks of networks built around specific business projects.

In BIPS there is not a political structure that sets minimal conditions of payment or labor conditions, for example, for those who can benefit the least of the facilities that mediums such as Internet provide. However, it must be reminded that not even classical political systems can guarantee bioethical labor conditions, despite the fact that many of them have specific legislation for it. In bio-inspired political systems, individuals become *their own force of work,* which implies more diversity in the market. Ultimately, diversity reduces the possibilities of monopolies to be established. Hence, the self-organization of economic structures in BIPS would not point toward savage economic models –like today´s. We hope that diversity in catalaxies entail that the influence of enterprises stops being feasible to be described by a power law.

**5.2 A Complex World with Visa Policies is Non-Viable**

Political regimes impose restrictions to international mobility by means of visa policies, international treats and agreements. Bio-inspired political systems would not forcibly try to maintain a population homogeneous, protecting it from e*xternal perturbations.* That is, keeping it away from integrating it with citizens of different nationalities. On the contrary, BIPS would contribute to the mobility of individuals among human groups.

Three possible conditions could come from the phenomenon of traveling freely *abroad.* First, there could be a homogenization of religious, ethical, economic, cultural or philosophical aspects in human social systems. Second, the diversity of the latter could increase as a result of combining different religions, economies, products, cultures, subcultures, traditions, philosophies, etc., just as it happens with mutation and recombination in biology [43]. Third, both cases could occur simultaneously: the world could homogenize in some aspects and would become increasingly diverse in some others.

We state that the third is the most probable scenario –one that will boost decentralization and deterritorialization and will lead to a general wellbeing for individuals along the world. As economies would not be grounded anymore in the existence of currencies, economic wellbeing would not strictly base on the

*opportunities and capabilities of choosing* [44]. That is, the wider is the network of degrees of freedom of a community, the better are the economic opportunities for individuals and groups. Being connected can be linked with economic wellbeing. The more connected a community is and can be, the better opportunities of economic growth it has. The more complex its network of citizens mobilization, the better its level of commodities and comfort because greater are the probabilities of profiting from increased diversity coming from catalaxies. Although it cannot be yet affirmed with 100% certainty that there is casual direction between network diversity, connectivity factors and economic wellbeing, "social network diversity seem to be at the very least a strong structural signature for the economic development of a community" [45]. This correspondence can be confirmed by comparing economic conditions of the nations that are more connected today with those where citizens can barely travel abroad.

An additional argument that must be pointed out about the relation between diversity and complexity is that much diversity, whereas it is a variation, a distinction of species or a form of configuration, is not always a desired condition because the system can become ineffcient or catastrophic [43]. On the other hand, diversity limited by some homogenization and interrelated with complex adaptive rules and other characteristics of complexity can produce robustness within a system. In this case, a variety of political system. Diversity changes the equilibrium of a system - which is dynamic in human social systems and this will bring robustness to the sociopolitical dynamics of the world because the more *emergent* diversity there is, the more complexity there exists as well [43].

**5.3 True Democracy is only possible in a Politically Complex –Anarchist- World**

Many of contemporary Political systems in western world are representative democracies with regimes structured by tree topologies. The idea of bio-inspired political systems sustains itself in the fact that there is not a real need of having centralized control mechanisms for organizing human social systems because complex systems tend to self-organize. A completely self-organized world is a world that has implemented third-order models. Third-order models are possible in anarchist contexts, based on the possibilities of democratic principles because we have to consider that decisions will continue to be taken -in complex network topologies-, which makes our model an anarchic-democratic world.

Anarchy means having no principle or authority. The word is composed by the greek words *an -ἀν-* (without) and *arche -ἀρχή-* (principle), but there is nothing in is original meaning that links it with the absence of order. On the contrary, the great theorists of anarchism [46, 47, 48, 49] described anarchy as a political system where order is synthetized bottom-up [50]. Anarchy is a social synergy rich-realm because it bases, mainly, in emergent cooperation networks in a dissipative self-organized complex system. Therefore, anarchy is compatible with networks of self-regulation [51], cooperation and consensus. On the other hand, the idea of democracy originated in the greeks, from the words *demos* -δῆμος- (people) and the word *kratos* –κράτος- (power). Then, whereas anarchy means the absence of government, democracy means that the power relies on the people. Together, both ideas are not incompatible.

Stating that bio-inspired political systems leads to an anarchic-democratic world is to state that people, *those who are today in the base of the pyramid,* will be their own governors; that is, that they will self-organize. An anarchic-democratic world is a world where complex networks of sociopolitical interactions dynamically self-organize human social systems. The interactions between this complex networks give rise to adaptive control mechanisms for every particular case. Anarchy is a consequence of BIPS because there are no individuals or groups directing or trying to organize human social systems. In respect to democracy, as there are no intermediates for the decision-making processes, every individual can be part of the conformation of sociopolitical dynamics, so we can refer to a truly democratic stadium of human history, despite the absence of coercive regimes that guide democratic actions. This dynamic is one of the most complex possibilities for living the political and politics. Going one step forward, we claim that the topology of a true democracy is the same as the topology of anarchy: complex network structures, which is the topology of third-order models for BIPS.

Unlike today, in an anarchic-democratic world there would be no discrimination among physical or virtual presence for problem-facing, decision-making and decision-implementation because the adscription of individuals to a defined geographical space will not matter for their sociopolitical interactions. Instead, this will be replaced by an interest or expression of a chosen collectivity or individuality. One feature that will make this phenomenon possible will be that there will be no regimes that concentrate the power and the rights over the decision-making processes. As a consequence, there will not exist centralized top-down power structures that some groups will try to bring down.

We intrinsically blame top-down power structures for the existence of political violence, guerrillas, insurrection, para-institutions, rebellion, strikes, civil violence, wars, and rebels in the form of academic embryos because the institutionalization of politics by top-down structures segregates groups that operate outside the defined boundaries of political regimes, which makes them try to bring down the establishment –by any means. Without having only one way for recognizing sociopolitical dynamics, there would be no establishment to dwindle, pervade, cooptate, change or eliminate with the use of coercive mechanisms.

In complex contexts, political adaptiveness is *much more* preferable than rigidity and sometimes citizen participation mechanisms are not flexible enough. A truly complex adaptive political system in a flexible environment can only be achieved in anarchic contexts where individuals, if they wish, can behave democratically. In the end, both, (direct) democracy and anarchy are expressions of politics as *politéia.*

### 5.4 A Complex World much Closer to Bio-Ethics

Since the industrial revolution, the resources on earth have been extracted in augment. This has caused a deterioration of the conditions for life on the planet and has killed and extinct thousands of species in such a brief period of time that it can hardly be said that extinctions obeyed to natural causes [52]. Classical political systems are responsible for great part of the harm committed towards our planet and the life that inhabits it: governors in turn sign international treaties and form internal policies about the extraction, production, deforestation and managing of the natural resources in their political parcels. They define the limits, or the absence of them, to every action that involves the national territory: the mining activities and the extraction of resources, human labor conditions, limits and permissions for hunting and frames for the economic activities of corporations and multinationals. Thus, they decide: scopes, limits and controls. That is why, in great part, classical political systems can be directly blamed for affecting negatively the conditions for life on earth, and life itself. However, blaming specifically governors is pointless because, anyway, independently of who is or are in turn concentrating political power, the structures of political regimes are suitable for expecting negative bioethical outcomes. Basically, the central node, hub or control core in the tree topology, basically, can take any decision it wants, even if it is not beneficial for the lives of some groups or species that inhabit the political grains or parcels in question. In sum, classical political systems are directly responsible for the reduction of bio-diversity on earth.

Bio-inspired political systems are a step closer towards a pacific interaction among individuals and human social systems because of their bottom-up synthetized dynamics. Biologically motivated models are the best alternative that humans can adopt in order to properly organize the biodiversity that is left. BIPS point towards the self-organization of complex networks that synthetize how the use of earth´s resources are organized, defining the extraction, mining, deforestation, reforestation and hunting activities, using only local information and following decisions taken with the support of metaheuristics and modeling and simulation, and implemented by means of complex topologies. The combination of the three in BIPS might stop or reverse what can still be pulled back. For avoiding the disastrous situations currently experienced by the planet, subsequently of BIPS there would be synergies coming from the non-linear interactions of these complex networks. It is hoped that the idea of synthetizing bottom-up political systems permeates the mainstream of science and the political, in order not to return any longer to the type of top-down political structures that have harmed so profoundly life on earth. With no doubt, BIPS imply a network comprehension of living systems´ relations and interactions, which entails bio-ethical ways for organizing social systems.

In addition, less antropocentric ways of organizing and structuring our human social systems will positively influence the way we exploit *Gaia*, evolving towards more dynamic equilibriums between human social systems, natural social systems and artificial systems. The road, however, is not yet paved. We still have to gain more knowledge about the complex computational dynamics of live, social systems and artificial life. Finding this type of adaptive balance is, with no doubdt, the greatest challenge ever faced in human history [53]. Bio-inpired political systems are the only way in which, if political systems comtinue to exist, a political system can harness the increasing complexity of the interactions in our world.

Figure 7 conjugates the mentioned backgrounds with the implications of Bio-inspired political systems.

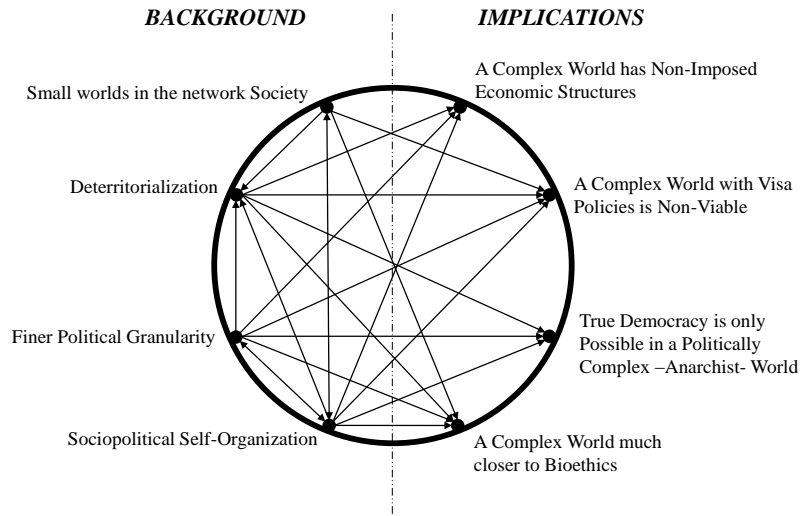

Fig. 7. Backgrounds and Implications

## 6 Concluding Remarks

We stated that bio-inspired models for political systems mean materializing complexity. Agent-based modeling and simulation helps gaining comprehension about the systems modeled and testing decisions before implementing them. Metaheuristics are useful for problem-solving and exploring spaces of solutions that lead to solutions close to the optimal. And complex models for topologies permit generating congruence between the structure of political systems, the types of problems they must solve and the complex systems upon which they decide.

The complexity of BIPS would provide them with more degrees of freedom than classical political systems, for modifications to occur in their structures and dynamics *in accordance* with the evolution of human social systems. Their adaptability, based on the composition of the system and the types of relations between the nodes, would give them the capacity of pacifically generating variations on every temporal control parameter as the *politéia* synthetizes them.

This paper highlights the necessity of complexifying the means by which human social systems organize (political systems), since there is still a vacuum in visualizing the profound political implications of standing in complexity for thinking the political. For our case, political systems, imagining them in contexts of complexity leads necessary to think about ruptures, discontinuity, decentralization, evolvability, transformations and, of course, self-organization, which is really lacking in the mainstream of the studies of the political. Indeed, most of the theorists of politics that work in complexity and are interested in the study political systems are still theorizing in terms of voting dynamics, elections and governability. They assume that complexity is not being constrained by classical institutionalization mechanisms because human social systems present complex behaviors. They also assume that political systems are already self-organized. [54]

Modeling and simulation added to metaheuristics plus complex topologies for political regimes can help us transform current political systems towards more organic ones, coping with the complexity of human sociopolitical dynamics. But they can also enlight us with an idea about the adjacent possible [55] of actual political systems. Based on this, the need of a field for the study of bio-inspired models for political systems is imperative. It is time for political science to underline the phenomenon that sooner or later will lead us to pass by the *social contract* era.

**Acknowledgments:** I sincerely thank Nelson Alfonso Gómez Cruz for the fructiferous conversations we had during the writing of this paper, his comments, suggestions and improvements on my graphics. I also acknowledge Carlos Eduardo Maldonado for his valuable corrections on the paper and for the one year internship I did with him, where I could work on developing my ideas. And I appreciate the precious time and suggestions coming from Beatriz Franco Cuervo.